# THz Superradiance from a GaAs: ErAs Quantum Dot Array at Room Temperature

W-D. Zhang,[1,2*] E. R. Brown,[1,2*] A. Mingardi,[1] R. P. Mirin[3], N. Jahed[4] and D. Saeedkia[4]

[1]Departments of Physics and Electrical Engineering Wright State University, Dayton, OH, USA 45435
[2]Terapico, LLC, Beavercreek, OH, USA 45431
[3]Applied Physics Division, National Institute of Standards and Technology, Boulder, CO, USA 80305
[4] TeTech S, Suite 3, 170 Columbia, W. Waterloo, Ontario, N2L 3L3 Canada
*Corresponding author: wzzhang@fastmail.fm, elliott.brown@wright.edu

*Abstract: We report experimental evidence that an ErAs quantum-dot array in a GaAs matrix under 1550 nm pulsed excitation produces cooperative spontaneous emission—Dicke superradiance—in the terahertz frequency region at room temperature.*

In 1954 Dicke proposed that cooperative spontaneous emission could occur among a collection of dipoles through interaction with a common electromagnetic radiation field[1]. He named the phenomenon "superradiance," which was subsequently observed in atomic/molecular systems[2], and in quantum dot systems[3]. To date, almost all of these systems are either extended media or require a pencil-shaped (i.e., high directivity) beam. In this letter, we present evidence of cooperative spontaneous emission from the dipoles of a quantum dot (QD) array in a volume smaller than or comparable to the wavelength of radiation.

The ErAs QD array in GaAs was obtained by growing a GaAs epitaxial layer heavily doped with erbium on a semi-insulating GaAs substrate by molecular beam epitaxy (MBE) (Fig. 1 (a)). The thickness of the epilayer was L=2 μm, and the Er doping ~8.8x10$^{20}$ cm$^{-3}$ such that the erbium incorporated into the GaAs in the form of ErAs quantum dots. This was proven by transmission-electron-microscope (TEM) imagery (Fig. 1 (b)). The most likely diameter is estimated to be~2.0 nm, and the density of quantum dots is $n_{QD}$~1×10$^{18}$/cm$^3$.

These ErAs quantum dots contain several quasi-bound levels near the mid-gap of GaAs, allowing a significant bound-to-bound absorption around 1550 nm. This is proven by the attenuation coefficient at 1550 nm $\alpha_{1550} \sim 7.4 \times 10^3$ cm$^{-1}$ derived from infrared transmission measurements. Photoconductive (PC) switch devices were fabricated by patterning a self-complementary square spiral on the top of the epilayer with standard image reversal processing (Fig. 1 (c)). The antenna has an active area at the center of 9 μm×9 μm, which is compatible with the spot size of a focused 1550-nm pump fiber mode-locked laser having 90 fs pulses and 100-MHz repetition frequency. A significant amount of THz power, ~117 μW average, was measured with a calibrated Golay cell detector at room temperature. Figure 1 (c) exhibits a curve of THz power vs. 1550 light intensity at a bias voltage of $V_b$=100 V. It is fit well by a power law: $P_{THz} \propto P_{1550}^{2.15}$. This is close to the Dicke theory, i.e. $P_{THz} \propto N^2$, where $N$ is the number of excited dipoles, since N is also proportional to the peak 1550 nm laser power ($N \propto P_{1550}$). Figure 1 (d) displays the power spectrum obtained after FFT of the interferogram obtained with a Michelson autocorrelator.

We measured the waveform of the THz pulses from the PC switch with a THz time-domain spectrometer. The switch was used as transmitter, pumped separately by 1550- and 780-nm sub-picosecond pulses. Commercial InGaAs (for 1550 nm) and LT:GaAs PC switches (for 780 nm) were used as the receivers. The 780 nm result is a benchmark since it is known to produce electron-hole pairs across the GaAs bandgap, utilizing the quantum dots primarily as recombination centers. Figure 2(a) shows the THz waveform for a 780 nm, average pump power of ≈14 mW. Only one pronounced THz peak appears in the waveform as expected, indicating the THz radiation is caused by the acceleration of the photocarriers in the large bias field along with their ultrafast recombination. This is typical behavior for cross-bandgap ultrafast photoconductive devices.

Figure 2(b) shows the THz waveform for the 1550 nm pump with an average power of ~23 mW. Unlike the 780-nm waveform, a damped ringing occurs over four cycles, with the period increasing as the waveform evolves. This ringing is a telltale feature of superradiance, which has been observed experimentally in several material systems [4], and studied theoretically[5]. It originates from the feedback between a radiation field and a *collective* system with multiple optical dipoles. The radiation field swings toward both positive and negative polarities, displaying a displacement-current-like alternation, which can't be explained as normal photoconductive relaxation. And the fact that the same device pumped at 780 nm displayed no ringing rules out other possible effects, such as resonances in the spiral antenna. Furthermore, to check for possible nonlinear-optical effects in the detector, or self-interference along the THz beam path, we replaced the QD PC switch device with a commercial $In_{0.53}Ga_{0.47}As$ cross-bandgap-excited PC switch with the same 1550 nm setup and the same receiver (Fig. 3). No such ringing was observed. Figure 2(c) shows the power spectrum of 2(b) after an FFT operation. Like Fig. 1(d), most of the power occurs in a band-limited range between ~100 GHz and 500 (1(d)) or 600 (2(c)) GHz.

The observed superradiance effect is consistent with theoretical investigations[6, 7] and experimental studies[3] that QDs in an ensemble —even an inhomogeneous one—can't be treated as isolated individuals when they are coupled to the same radiative field in a volume having dimension << λ. The interaction between the energy levels of the QDs and their spontaneous emission field can cause cooperation among the QDs, which makes the emission become superradiant. Our modeling treats the quantum dots as spherical wells with the wavefunctions having a hybrid character–the usual labeling by principal and angular-momentum quantum numbers as in any central potential, but also a band labeling from the crystalline (rocksalt) structure of the ErAs. In the typical ~2.0-nm dots in Fig. 1(b), both effects are important as is excitonic binding, which is known to be much stronger in quantum dots than in bulk crystals, and can display a "giant oscillator strength" [8] that favors superradiant behavior.

The total number of QDs under the direct illumination of the pump pulses is estimated by $N \approx n_{QD}W^2L = 1.62 \times 10^8$ (Fig. 1 (a)). The polarization is the summation over all available dipoles, and $\mathbf{P} = \sum_i^N \mathbf{p}_i \approx N\mathbf{p}$ when the dipoles become aligned by the strong ultrafast laser field. The initial tipping angle of the Bloch vector **P** is approximated by $\theta_0 \approx 1/\sqrt{N} = 7.9 \times 10^{-5}$.[5] Assuming the dipoles radiate coherently, we solved the sine-Gordon equation with this initial value and obtained the "ringing" electromagnetic radiation field plotted in Fig. 2(d). Only one other quantity was required for the solution, the superradiance time constant $T_R$ of ≈201 fs. And we estimated the individual quantum-dot spontaneous lifetime $T_{sp}$ to be ≈10 ns. The fact that, $T_R << T_{sp}$ is another characteristic of superradiant systems.

Figure 4 shows how THz power is radiated from the square-spiral antenna. The time varying, macroscopic polarization **P** corresponds to a displacement current. And through Maxwell's dynamic equation and current continuity, this induces a proportionate conduction-current density into the metallic arms of the antenna.

**Funding.** Army Research Office (W911NF-12-1-0496 and SBIR W911NF-17-P-0069).

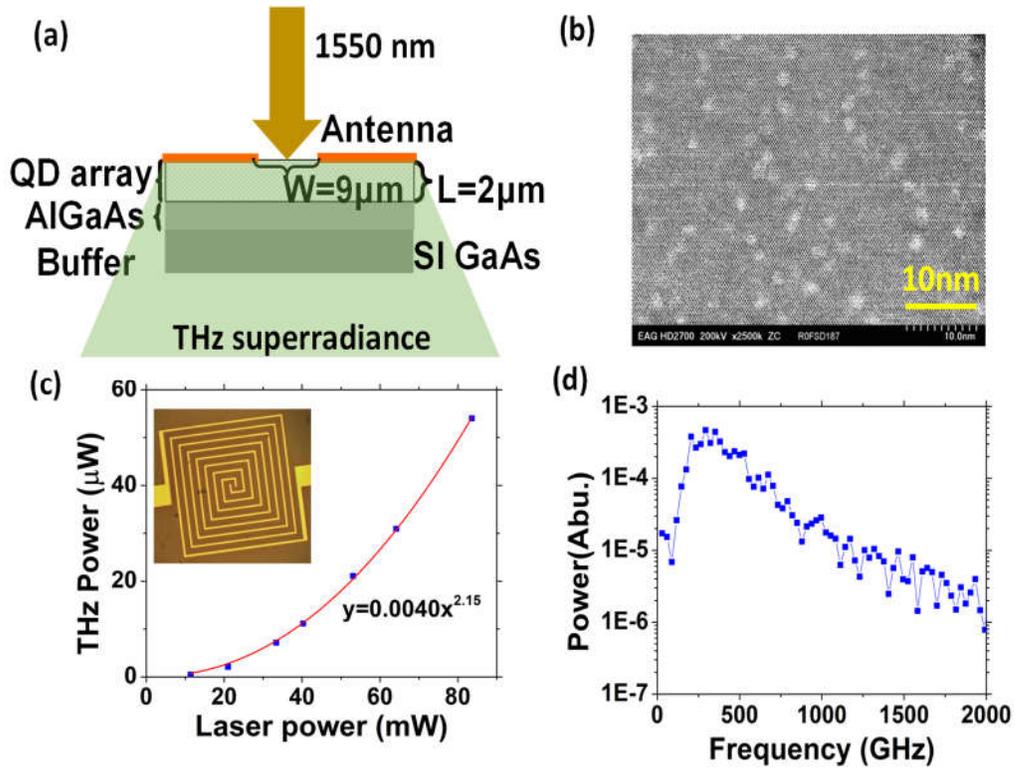

**Fig. 1**. (a). The side view of an ErAs quantum dot array embedded in a GaAs epilayer. (b) a TEM image (cross-sectional review). (c)The THz power vs. the1550 laser power. (d) The power spectrum analyzed with an autocorrelator.

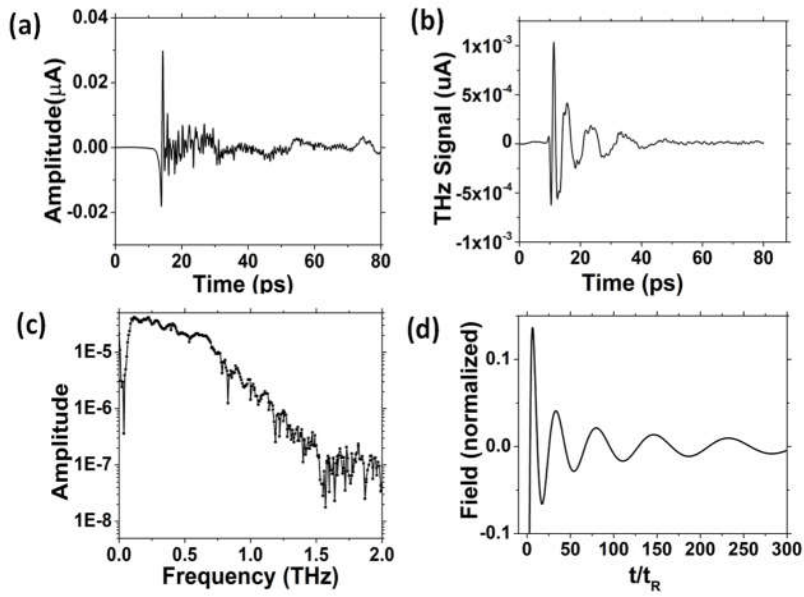

**Fig. 2.** The waveform with 780-nm pump (a), and 1550-nm pump. (b). (c) The FFT spectrum of the 1550 nm waveform. (d) Simulated "ringing" pattern produced with the sine-Gordon equation.

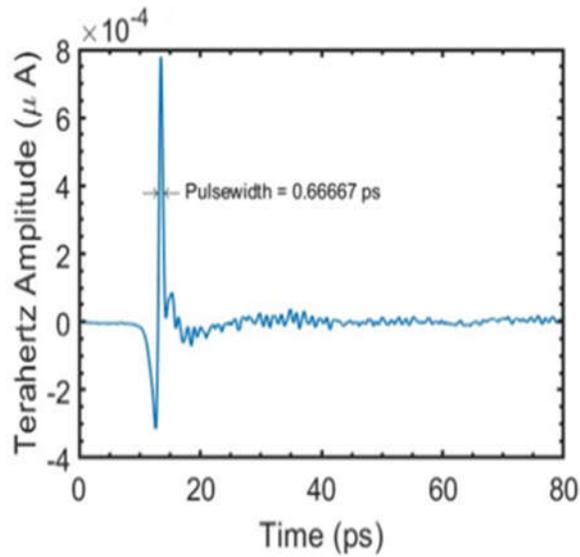

**Fig. 3.** The waveform of an InGaAs PC device with 1550 nm pump.

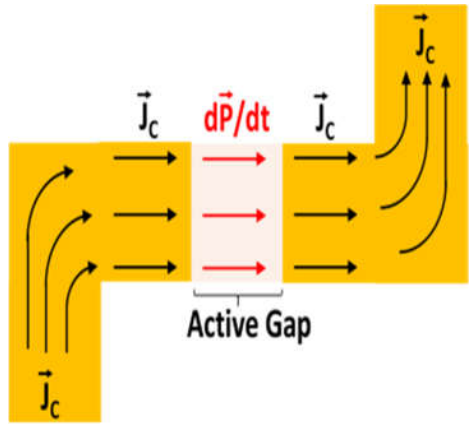

**Fig. 4.** Coupling between polarization current in gap and conduction current in arms of spiral antenna.